\documentclass[aps,prb, twocolumn,showpacs,preprintnumbers,superscriptaddress]{revtex4}

\usepackage[dvipdfm]{graphicx}% Include figure files
\usepackage{dcolumn}% Align table columns on decimal point
\usepackage{bm}% bold math

\begin{document}

\title{Effect of magnetic field on the superconducting phase  in the electron-doped  metallic double-chain compound Pr$_{2}$Ba$_{4}$Cu$_{7}$O$_{15-\delta }$}

\author{Taiji Chiba}
\author{Michiaki Matsukawa} 
\email{matsukawa@iwate-u.ac.jp }
%\thanks{}
\author{Junki Tada} 
\author{Satoru Kobayashi} 
\affiliation{Department of Materials Science and Engineering, Iwate University, Morioka 020-8551, Japan}
\author{Makoto Hagiwara}
\author{Tsuyoshi Miyazaki}
\affiliation{Kyoto Institute of Technology, Kyoto 606-8585,Japan}
\author{Kazuhiro Sano}
\affiliation{Department of Physics Engineering, Mie University, Tsu 514-8507,Japan}
\author{Yoshiaki $\bar \mathrm{O}$no}
\affiliation{Department of Physics, Niigata University, Niigata 950-2181,Japan}
\author{Takahiko Sasaki}
\affiliation{Institute for Materials Research, Tohoku University, Sendai 980-8577,Japan}
\author{Jun-ichi Echigoya$^{1}$}

\date{\today}

\begin{abstract}
We report the magnetotransport and $dc$ magnetic susceptibility  of the polycrystalline samples of Pr$_{2}$Ba$_{4}$Cu$_{7}$O$_{15-\delta }$, to examine the effect of magnetic field on the superconducting phase of the metallic CuO double chain.  
%A reduction treatment of the as-sintered sample in vacuum causes higher superconductivity achieving $T_{c,on}=\sim 30$ K for $\delta =0.94$.  
The resistive critical magnetic field is estimated to be about 21 T at low temperatures from the resistive transition data. On the other hand, the corresponding critical field determined from the magnetization measurements gives rise to a very low value of $\sim 0.3$ T at 2 K. These discrepancies in the magnetic response between the resistivity and magnetization data are caused  by  disappearance  of the magnetically shielding effect even in relatively lower fields. In spite of the observation of the resistive drop associated with the superconducting transport currents, the suppression of the diamagnetic signal is probably related to the superconductivity of quasi one-dimensional CuO double-chain.  The behavior of Seebeck coefficient in the superconducting Pr$_{2}$Ba$_{4}$Cu$_{7}$O$_{15-\delta }$ is discussed on the basis of the double chain model from the density functional band calculation.

%Seebeck coefficient of the superconducting sample shows a metallic conduction, followed by  a clear drop below $T_{c,on}$ and is in its temperature dependence below 100 K quite different from that of the non-superconducting one. This finding strongly suggests a dramatic change of the electronic state along the CuO double chain due to the reduction treatment for the appearance of superconductivity .
\end{abstract}

%\keywords{}
\renewcommand{\figurename}{Fig.}
\maketitle

\section{INTRODUCTION}
\begin{figure}[ht]
\includegraphics[width=6cm]{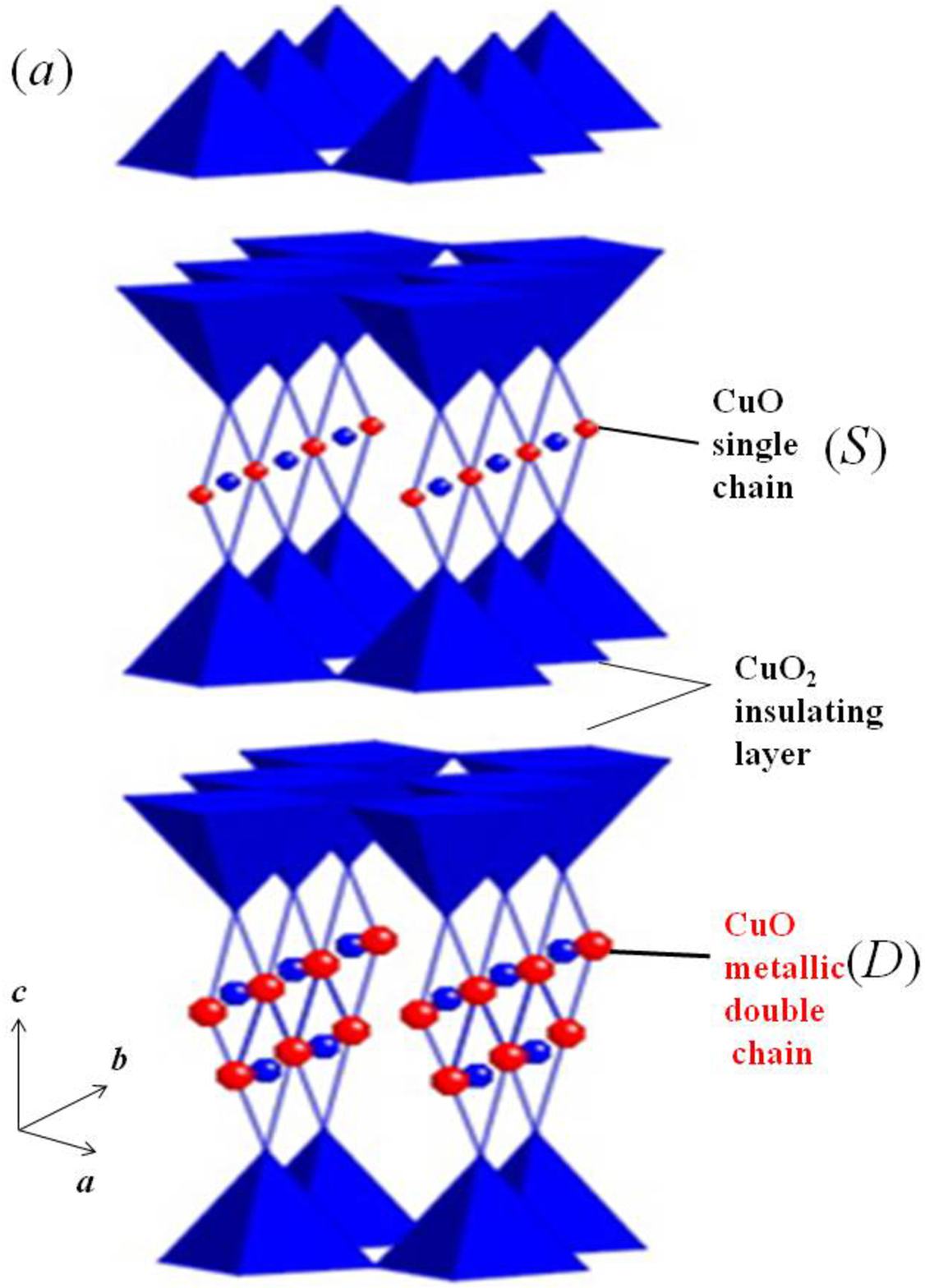}
\includegraphics[width=8cm]{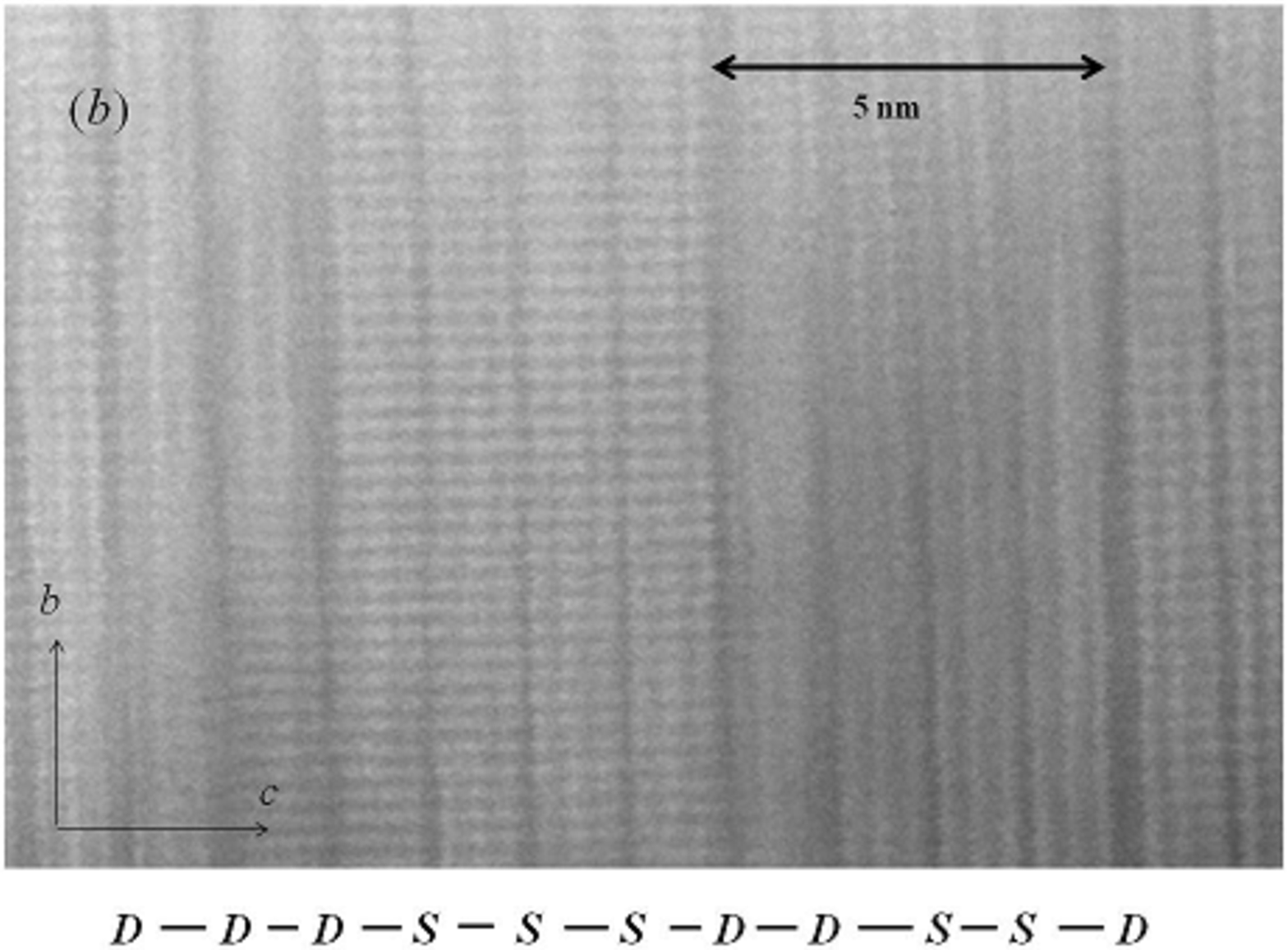}
\caption{(color online) (a)Typical crystal structure of the electron-doped  metallic double-chain compound Pr$_{2}$Ba$_{4}$Cu$_{7}$O$_{15-\delta }$. The CuO double chain structure is parallel to the direction of $b$-axis. (b) TEM image of the high-$T_\mathrm{c}$ sample exhibiting $T_\mathrm{c}^\mathrm{on}=30.5$ K prepared by a sol-gel technique\cite{TO12}. $S$ and $D$  denote CuO single chain and double chain blocks.
% (c) The x-ray diffraction pattern of the  Pr$_{2}$Ba$_{4}$Cu$_{7}$O$_{15-\delta }$ with $T_{c,on}=26.5$ K synthesized by a citrate pyrolysis method. 
 }\label{TEM}
\end{figure}
\begin{figure}[ht]

\includegraphics[width=8cm]{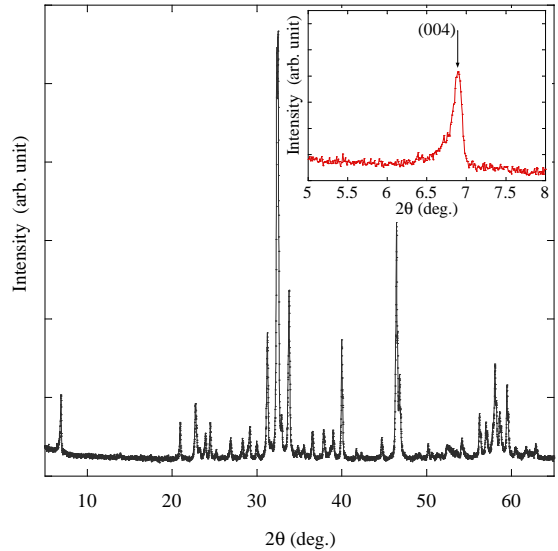}
\caption{(color online) The x-ray diffraction pattern of the  Pr$_{2}$Ba$_{4}$Cu$_{7}$O$_{15-\delta }$ with $T_\mathrm{c}^\mathrm{on}=26.5$ K synthesized by a citrate pyrolysis method. The inset represents the low angle diffraction data enlarged near $2\theta =5^{\circ }-8^{\circ }$. The allow points to the peak corresponding to the Miller index (004) of Pr247. Neither peaks of Pr123 nor Pr124 were visible. 
 }\label{Xray}
\end{figure}
\begin{figure}[ht]
\includegraphics[width=9cm]{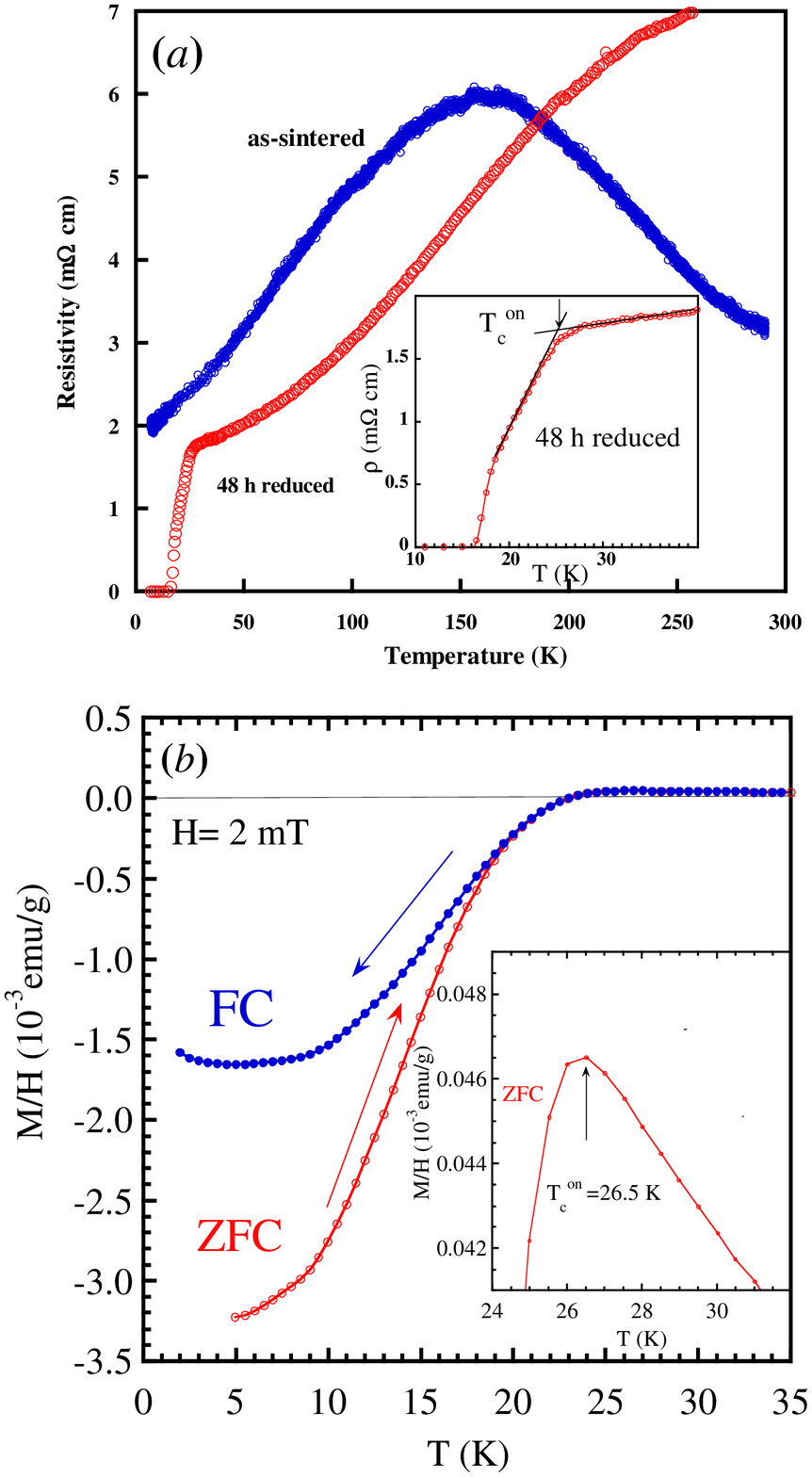}
\caption{(color online) (a) Temperature variation of the electrical resistivity for the Pr$_{2}$Ba$_{4}$Cu$_{7}$O$_{15-\delta }$ sample  reduced for 48 h in vacuum. For comparison, the data of the as-sintered sample are shown.
The inset displays the enlarged  data of the reduced sample around $T_\mathrm{c}$. (b) Temperature dependence of the magnetic susceptibility M/H of the reduced Pr$_{2}$Ba$_{4}$Cu$_{7}$O$_{15-\delta }$ measured in a magnetic field of 2 mT after zero-field and field cooling (ZFC and FC). In the inset, the magnified data are plotted to clarify the definition of $T_\mathrm{c}^\mathrm{on}$. 
 }\label{RT}
\end{figure}

Since the discovery of high-$T_\mathrm{c}$ copper-oxide superconductors,  extensive studies 
on strongly correlated electron system  have been in progress on the basis of physical properties 
of two-dimensional (2D) CuO$_{2}$ planes.  Moreover, from the viewpoint of low-dimensional physics, 
particular attention is paid to the physical role of one-dimensional (1D) CuO chains included 
in some families of high-$T_\mathrm{c}$ copper oxides.
% such as Y-based superconductors with the transition temperature $T_{c}$=$\sim$ 92K .  
It  is well known that the Pr-substitution for  Y-sites in YBa$_{2}$Cu$_{3}$O$_{7-\delta}$ (Y123) 
and  YBa$_{2}$Cu$_{4}$O$_{8}$ (Y124) compounds dramatically suppresses $T_\mathrm{c}$  and 
superconductivity in CuO$_{2}$ planes disappears beyond the critical value of  Pr , $x_\mathrm{c}$=0.5 and 0.8, 
respectively. \cite{SO87,HO98} Such a suppression effect due to Pr-substitution on superconductivity 
has been explained in terms of the hybridization model with respect to 
Pr-4$f$ and O-2$p$ orbitals.\cite{FE93}  
%Y124 compound with double chains is thermally stable up to 800 $^\circ$C ,while in Y123/7-$\delta$ oxygen deficiencies are easily introduced at lower annealing temperatures .\cite{JO87}  
Intermediate between PrBa$_{2}$Cu$_{3}$O$_{7-\delta}$ 
(Pr123) with CuO single chains and PrBa$_{2}$Cu$_{4}$O$_{8}$
(Pr124) with CuO double chains   is  the Pr$_{2}$Ba$_{4}$Cu$_{7}$O$_{15-\delta}$ (Pr247) compound 
with an alternative repetition of the CuO single chain and double chain blocks along the $c$-axis,
isostructural with superconducting Y$_{2}$Ba$_{4}$Cu$_{7}$O$_{15-\delta}$ (Y247). \cite{BO88,YA94} 
%The crystal structure of  orthorhombic Pr247 is schematically shown in Fig. 1 (space group $Ammm$).
%In contrast to Y123, Y247 retains its orthorhombic structure and its superconducting properties, even after oxygen along the single chains is completely depleted.\cite{GE92} This finding is considered to be due to the presence of the double CuO chain block in Y247. 
For the Pr247 compound, it is possible to examine physical properties of the metallic CuO double chains 
by controlling oxygen content $\delta  $  along the semiconducting CuO single chains.
From the anisotropic resistivity measurement of single crystal Pr124, we note that  its metallic transport is responsible for the conduction in the CuO double chains.\cite{HO00}

Matsukawa et al., discovered that oxygen removed polycrystalline Pr$_{2}$Ba$_{4}$Cu$_{7}$O$_{15-\delta }$   with  metallic CuO double chain  shows superconductivity around the onset $T_\mathrm{c}^\mathrm{on}=\sim 15$ K. \cite{MA04} 
In the Hall coefficient measurement of the superconducting Pr247, it is found that the main carrier is varied from hole to electron at intermediate temperatures below 120 K upon decreasing temperatures, indicating an electron doped superconductor.\cite{MA07}
A nuclear quadrapole resonance experiment conformed that the superconductivity of the reduced Pr247 is realized at the CuO double chains.\cite{WA05} 
For the appearance of Pr247 superconductor, it is essential to remove oxygen from the as sintered sample of Pr247 through reduction procedure and then electrons are probably doped to the CuO double chain layeres.\cite{MA04,YA05,HA07} 
%Through several works on synthesis of higher $ T_\mathrm{c}$ samples in nominal Pr247 composition, it has been made clear that the heterogeneous structure containing both Pr123 and Pr124 phases in Pr247 system plays a crucial role on attaining higher $ T_\mathrm{c}$. \cite{HA07,HA08} 
High-resolution transmission electron microscopy (TEM) analyses of the heterogeneous Pr247 sample  exhibits that there exists an irregular long-period stacking structure along the $c$ axis such as \{D-S-S-S-S-D\} sequence, where S and D denote CuO single chain and CuO double chain blocks, respectively.\cite{HA07}
Recently, Toshima et al., reported the effect of pressure on the magnetization, and transport properties in the nominal composition Pr$_{2}$Ba$_{4}$Cu$_{7}$O$_{15-\delta }$ synthesized by a sol-gel technique. \cite{TO12}
Their work reveals that the higher superconductivity of $T_\mathrm{c}^\mathrm{on}=\sim 30$ K for the $\delta =0.94$ sample is achieved  and is enhanced  up to 36 K under the applied pressure of  1.2 GPa.  
%In this paper, we have reported the magnetic, electronic and thermal transport properties of the oxygen removed  Pr$_{2}$Ba$_{4}$Cu$_{7}$O$_{15-\delta }$. In particular, the  $\delta =0.94$ sample  exhibits higher superconductivity attaining $T_{c,on}=\sim 30$ K. 
% reduction treatment of the as-sintered sample in vacuum causes the bulk superconductivity.  
%Application of hydrostatic pressure on the superconductive sample  enhances $T_{c,on}$ up to 35 K at 0.8 GPa, accompanied by a gradual suppression down to 32 K at 1.2 GPa. 
%Seebeck coefficient exhibits a metallic conduction  below 200 K, followed by a substantial drop near $T_{c,on}=\sim 30$ K.
From the theoretical point of view,  the superconductivity of Pr247 was clarified by using the CuO double chain model.\cite{SA05,NA07,OK07,BE07} Sano et al., propose an electronic phase diagram including the superconducting phase in the weak coupling limit on the basis of the Tomonaga-Luttinger liquid. The doping dependence of superconductivity obtained in ref.\cite{NA07,HA11} is qualitatively consistent with the experimental results in Pr247.

In this paper, we demonstrate the effect of magnetic field on the superconducting phase  in the electron-doped metallic double-chain compound Pr$_{2}$Ba$_{4}$Cu$_{7}$O$_{15-\delta }$ with higher $T_\mathrm{c}^\mathrm{on}=26.5$ K. 
In Fig.\ref{TEM}, let us display a typical crystal structure of Pr$_{2}$Ba$_{4}$Cu$_{7}$O$_{15-\delta }$, where the CuO metallic double chain and semiconducting single chain stack alternatively along the $c$-axis. 
In the next section, the experimental outline is described. In Sec. III, we show the magnetotransport and $dc$ magnetic susceptibility  of the polycrystalline samples of Pr247 with a magnetic phase diagram. The temperature dependence of Seebeck coefficient in the as-sintered and reduced Pr247 is given with the magneto thermal transport data and is discussed on the basis of the density of state calculated by the double chain model.  
The final section is devoted to the summary. 

\begin{figure}[ht]
\includegraphics[width=9cm]{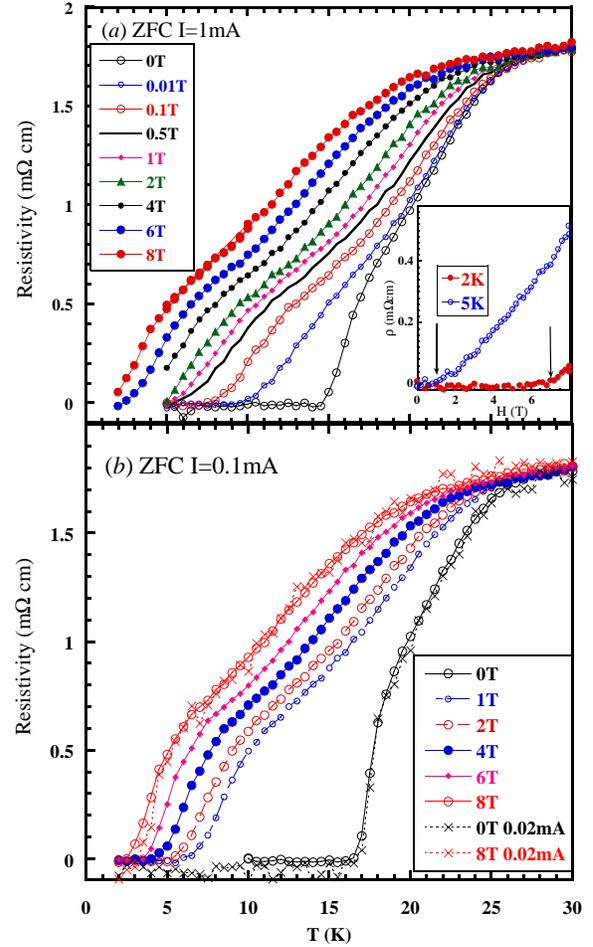}
\caption{(color online) (a) Temperature variation of the electrical resistivity in various magnetic fields $H=0,0.01,0.1,0.5,1,2,4,6$, and 8 T for the reduced sample of Pr$_{2}$Ba$_{4}$Cu$_{7}$O$_{15-\delta }$. The applied current $I=$1 mA is perpendicular to the magnetic field $H$ ($H\perp I$). The data are collected upon warming $T$ after zero field cooling. The inset displays the magnetoresistance effect as a function of magnetic field up to 8 T both at 2 and 5 K, where the allows denote $H_\mathrm{c}^\mathrm{zero}$. (see the caption of Fig.\ref{PD})
In the case of $I=$0.1 mA in $H=0,1,2,4,6$, and 8 T, the resistivity data are presented in (b). For comparison, the resistivity data taken at $I=$0.02 mA are added both in 0 and 8 T. 
 }\label{RTH}
\end{figure}
\begin{figure}[ht]
\includegraphics[width=9cm]{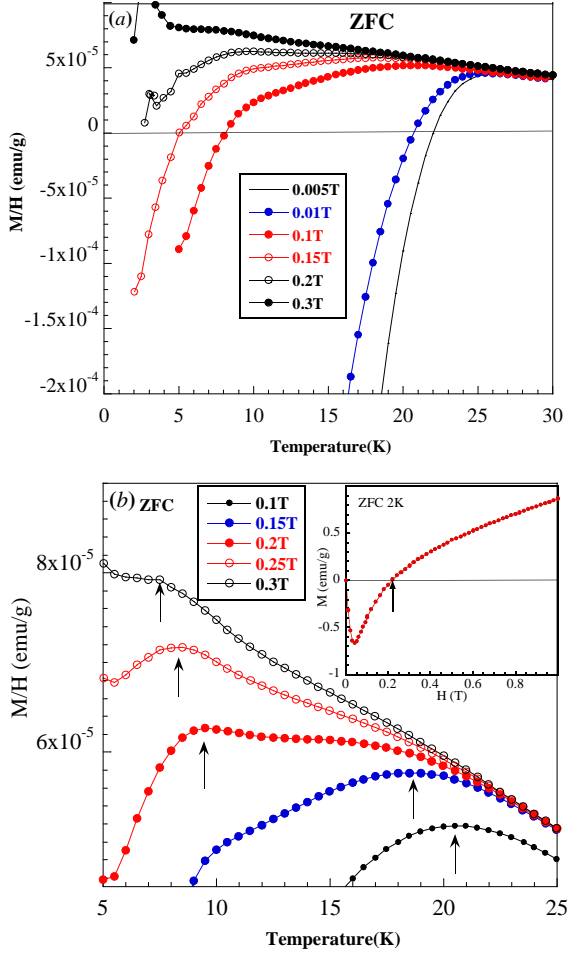}
\caption{(color online) (a) Temperature dependence of the $dc$ magnetic susceptibility in $H=0.005, 0.01, 0.1, 0.15, 0.2$, and 0.3 T for the reduced sample of Pr$_{2}$Ba$_{4}$Cu$_{7}$O$_{15-\delta }$ after zero field cooling. In (b), its magnified data are presented. The arrows denote the superconducting transition onset temperature $T_\mathrm{c}^\mathrm{on}$.  After ZFC, the field dependence of the isothermal magnetization  up to 1 T at 2 K is displayed in the inset of (b). 
 }\label{MTH}
\end{figure}

\begin{figure}[ht]
\includegraphics[width=9cm]{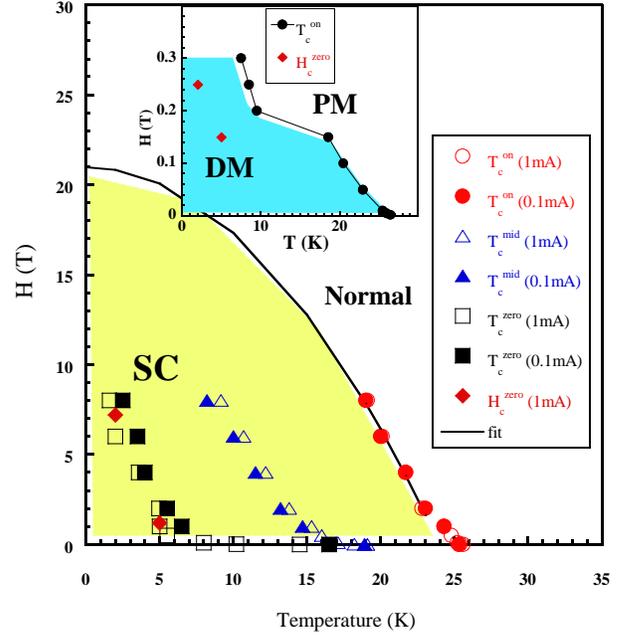}
\caption{(color online) Temperature dependence of the superconducting(SC) critical field $H_\mathrm{c}$($T$) of the reduced sample of Pr$_{2}$Ba$_{4}$Cu$_{7}$O$_{15-\delta }$. The onset $T_\mathrm{c}$ ($T_\mathrm{c}^\mathrm{on}$) and the midpoint $T_\mathrm{c}$ ($T_\mathrm{c}^\mathrm{mid}$) are determined from the resistive data as described in the text. The zero-point $T_\mathrm{c}$ ($T_\mathrm{c}^\mathrm{zero}$) denotes the value of the critical temperature  reaching  the zero-resistance state. The value of $H_\mathrm{c}^\mathrm{zero}$ is defined from the field sweep data  as the magnetic field where the zero-resistance state is violated upon increasing the field.  
For comparison, the inset shows the phase diagram separating between the diamagnetic(DM) and paramagnetic (PM) phases established from the $dc$ magnetic susceptibility measurement in Fig.\ref{MTH}. $H_\mathrm{c}^\mathrm{zero}$ means the magnetic field where the diamagnetic signal vanishes upon increasing the field at fixed temperatures. 
}\label{PD}
\end{figure}
\begin{figure}[ht]
\includegraphics[width=9cm]{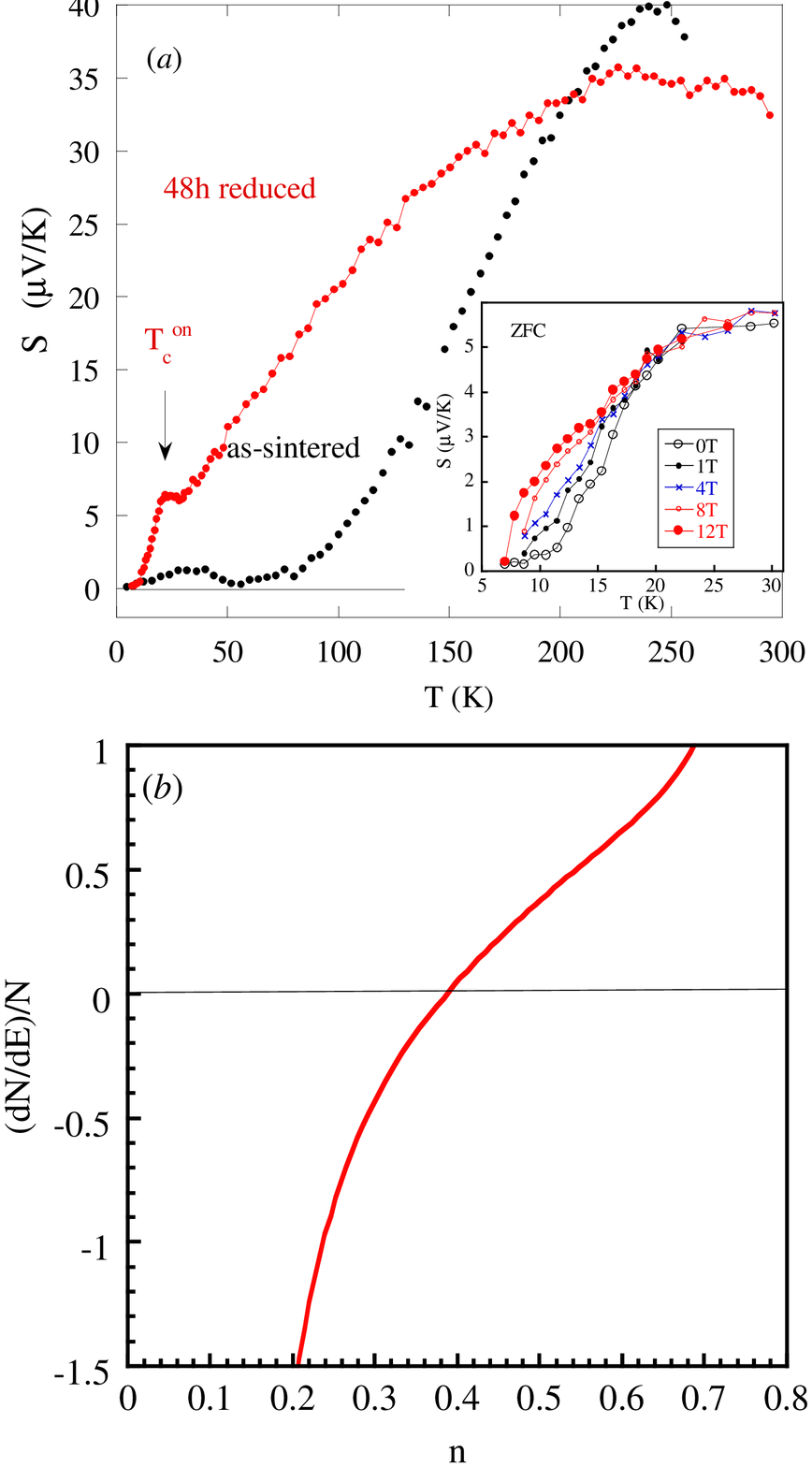}
\caption{(color online)(a)Temperature dependence of Seebeck coefficient for the reduced and as-sintered samples of Pr$_{2}$Ba$_{4}$Cu$_{7}$O$_{15-\delta }$. In the inset, Seebeck coefficient data of the reduced sample at lower temperatures are given under several magnetic fields of 0, 1, 4, 8, and 12 T after zero field cooling. 
 (b) The energy derivative of the DOS, $\frac{1}{N}\frac{dN}{dE}$, as a function of the electron density $n$. 
 The parameters used in this calculation are listed in ref.\cite{HA11}.  
}\label{S}
\end{figure}

\section{EXPERIMENT}
Polycrystalline samples of Pr$_{2}$Ba$_{4}$Cu$_{7}$O$_{15-\delta }$(Pr247) were synthesized using citrate pyrolysis and sol-gel methods.\cite{TO12,HA06} After several annealing processes,  resultant precursors were pressed into  a pellet and  it was then calcined at 875$ \sim $887 $ ^{ \circ }$C for a long time over 120-180 h under ambient oxygen pressure.
The reduction treatment on its as-sintered sample in vacuum  at 500 $ ^{ \circ }$C for 48 h yields oxygen removal, resulting in a realization of the superconductivity. 
A typical dimension of the pelletized rectangular sample is $9.9\times 2.4\times 1.7$ mm$^{3}$.
High-resolution transmission electron microscopy of the Pr247 sample with $T_\mathrm{c}^\mathrm{on}=30.5$ K was observed with an FEI-Titan microscope operated at 300 kV at Tohoku University.
Experimental images were obtained by using the high-angle annual drak-field scanning transmission electron microscope (HAADF-STEM) method for local crystal structure analysis.\cite{PE90} 
The oxygen deficiency  estimated from gravimetric analysis attain  $\delta = 0.56$ for the sample with $T_\mathrm{c}^\mathrm{on}=26.5$ K  prepared by the citrate method. For the sample synthesized by sol-gel method,  $\delta$ is 0.96 for 72 h reduction, giving $T_\mathrm{c}^\mathrm{on}=30.5$ K. 
The electric resistivity in zero field was measured with $dc$ four terminal method. The measurement of magnetotransport was carried out with the $ac$ four-probe method using the Physical Property Measuring System (Quantum Design, PPMS) from 2 K up to 40 K in zero field cooling. The electric current is applied to the longitudinal direction of the sample, ranging from 0.02 mA up to 1 mA. 
The applied magnetic field $H$ is perpendicular to the direction of the applied electric current $I$ ($H\perp I$).  Seebeck coefficient $S(T)$ was determined from both measurements of a thermoelectric voltage and temperature difference along the longitudinal direction of the measured sample. The magnetic field $H$ perpendicular to the   thermal current $Q$ ($H\perp Q$) was applied up to 12 T in a superconducting magnet at the High Field Laboratory for Superconducting Materials, Institute for Materials Research, Tohoku University.  
 The Seebeck coefficient of copper lead was subtracted from the measured values using the published data. 
The thermal conductivity data were collected with a conventional heat flow method.  
The $dc$ magnetization was performed under the zero field cooling (ZFC) process of 0.002$\sim  $1 T using the commercial superconducting quantum interference  device (SQUID) magnetometer (Quantum Design, MPMS). 

\section{RESULTS AND DISCUSSION}

First of all, let us show in Fig.\ref{TEM}(b), the TEM image of the high $T_\mathrm{c}$ sample with $T_\mathrm{c}^\mathrm{on}=30.5$ K obtained by the HAADF-STEM method.\cite{PE90}  We find out that there exists an irregular long-period stacking structure along the $c$ axis such as \{-D-D-D-S-S-S-D-D-S-S\} sequence.\cite{HA07}  Here, $S$ and $D$ at the bottom of (b), denote CuO single chain and double chain blocks along $b$-axis. This finding is close to higher carrier concentration with $\delta = 0.94$, resulting in higher $T_\mathrm{c}$, in comparison to the regular stacking case of \{-D-S-D-S-D\}.
Figure \ref{Xray} shows the x-ray diffraction pattern of the  Pr$_{2}$Ba$_{4}$Cu$_{7}$O$_{15-\delta }$ with $T_\mathrm{c}^\mathrm{on}=26.5$ K synthesized by a citrate pyrolysis method. The inset represents the low angle diffraction data enlarged near $2\theta =5^{\circ }-8^{\circ }$. The allow points to the peak corresponding to the Miller index (004) of Pr247. No peak of Pr123 or Pr124 phase was detected except for a small amount of BaCuO$_{2}$ around $2\theta =28^{\circ }-30^{\circ }$.   

Now, the temperature variation of the electrical resistivity for the 48 h reduced sample in vacuum  is shown in Fig.\ref{RT} (a). For comparison, the data of the as-sintered sample are plotted. The reduction treatment in vacuum results in the appearance of superconducting state with $T_\mathrm{c}^\mathrm{on}$=26.5 K. Furthermore, to check the bulk superconductivity, let us measure the magnetic susceptibility M/H of the reduced Pr$_{2}$Ba$_{4}$Cu$_{7}$O$_{15-\delta }$ measured in a magnetic field of 2 mT after zero-field and field cooling. Figure \ref{RT} (b) exhibits a diamagnetic signal below $T_\mathrm{c}^\mathrm{on}$=26.5 K both in the ZFC and FC curves, which is in good agreement with the resistive onset temperature. In particular, the superconducting volume fraction due to the shielding effect is estimated to be $\sim $30 $\%$ from the ZFC value at 5 K, indicating the appearance of the bulk superconductivity.   

Next, we try to measure the magnetotransport property of the present sample, to examine the magnetic effect on the superconducting phase of the metallic double chain.
 The resistivity starts to decrease below $T_\mathrm{c}^\mathrm{on}$=26.5 K,then follows a substantial drop, and finally achieves a zero-resistance state at $T_\mathrm{c}^\mathrm{zero}=\sim $15 K. The behavior of the resistive transition in zero field is changed to a widely broad resistive transition at the presence of magnetic field as shown in Fig. \ref{RTH}(a).  
At the maximum field of 8 T, we observe $T_\mathrm{c}^\mathrm{on}$=20.0 K from 26.5 K at 0 T , and  the value of $T_\mathrm{c}^\mathrm{zero}$ is considerably lowed down to 2 K at 8 T from 15 K at 0 T. Further, to check the motion of magnetic flux due to the Lorenz force, we examine the behavior of the resistive transition under magnetic fields at the lower current $I=$0.1 mA.(Fig. \ref{RTH}(b)) It seems that there exists no large differences in the magnetic field effect on the resistive drop between 1 mA and 0.1 mA. In Fig. \ref{RTH}(b), we present the resistive transition  data of Pr$_{2}$Ba$_{4}$Cu$_{7}$O$_{15-\delta }$ taken at the lowest current of 0.02 mA.  In the applied field of 8 T, we notice a small variation of the value of $T_\mathrm{c}^\mathrm{zero}$ from 2 K at 1 mA through 2.5 K at 0.1 mA to 3 K at 0.02 mA, indicating that these resistive drops are not caused by the motion of flux. 
%The superconducting critical field at 0 K is estimated to be 21 T from the $T_{c,on}$ versus $H$.
Figure \ref{MTH} displays the temperature dependence of dc magnetic susceptibility in $H=0.005, 0.01, 0.1, 0.15, 0.2$, and 0.3 T for the identical sample of Pr$_{2}$Ba$_{4}$Cu$_{7}$O$_{15-\delta }$ as used in the resistive measurement. 
The arrows in Fig. \ref{MTH}(b) represent the superconducting transition onset temperature $T_\mathrm{c}^\mathrm{on}$. From the temperature dependence of $M/H$,  the diamagnetic signal almost disappears at relatively lower values above $ \sim $0.3 T. The isothermal magnetization curve measured at 2 K in the inset of Fig.\ref{MTH}(b) is changed from its negative towards positive value across  almost 0.2 T.  
 This finding seems to be in disagreement with the resistive transition data reported here. This discrepancy in the magnetic response between the resistivity and magnetic susceptibility is caused by a strong suppression of the magnetically shielding effect in the latter measurement.  
Here, we evaluate from the resistive data taken at $I$=0.1 and 1.0 mA the temperature dependence of the superconducting critical field, to establish the superconducting phase diagram.  
The onset $T_\mathrm{c}$ ($T_\mathrm{c}^\mathrm{on}$) is determined from the intersection of the two lines  extrapolated  from the normal-state and superconducting-state resistivity, just above and below $T_\mathrm{c}$, respectively. The midpoint $T_\mathrm{c}$ ($T_\mathrm{c}^\mathrm{mid}$) is defined at the temperature where the resistivity decreases by a half of its value at $T_\mathrm{c}^\mathrm{on}$.  The values of $T_\mathrm{c}^\mathrm{on}$, $T_\mathrm{c}^\mathrm{mid}$, and $T_\mathrm{c}^\mathrm{zero}$ are determined from the temperature sweep data, while the value of $H_\mathrm{c}^\mathrm{zero}$ is defined from the field sweep data. 
The critical field at 0 K  is estimated to be $\sim $21 T from fitting the data of the onset $T_\mathrm{c}$ using the parabolic function formula of the critical field.  
For comparison, let us show in the inset of Fig. \ref{PD} the phase diagram separating between the diamagnetic and paramagnetic phases established from the $dc$ magnetic susceptibility measurement in Fig.\ref{MTH}. 
As mentioned before, the resistive critical field is, by about two orders of magnitude, much grater than the value of the critical field estimated from the magnetization data, which is closely related to the superconducting state of the CuO double chain. 
Following several theoretical works of a 1D interacting electron gas, it is well known that a 3D long-range ordering phase can exist only due to the finite interchain coupling.  The superconducting transition temperature, where 3D ordering of superconducting state is established in a quasi 1D system, is derived from the quantity $T_\mathrm{c}\sim t_{\perp }^{2}/t_{\parallel }$.\cite{SC83} Here, $t_{\parallel }$ and $t_{\perp }$ stand for intrachain and interchain transfer integrals, respectively. For a quasi 1D system, we point out a crucial role of coupling between chains due to Josephson tunneling when the 3D superconducting state is formed upon decreasing $T$.  Thus, the upper limit of $T_\mathrm{c}$ is estimated to be 29 K using  a typical value of $t_{b}\sim $0.5 eV and $t_{b}/t_{a}\sim  $14, where $t_{b}$ is the transfer integral along the $b$ axis of the CuO double chain and $t_{b}/t_{a}$ is the ratio of the transfer integral along the double chain to that across the chains.  The value of $t_{b}$ is taken from ref.\cite{HA11} 
and the ratio of $t_{b}/t_{a}$\cite{HO00} at low temperatures for single crystalline PrBa$_{2}$Cu$_{4}$O$_{8}$ is utilized as the experimental parameter. 

Finally, let us show in Fig.\ref{S} temperature dependence of Seebeck coefficient $S(T)$ of the polycrystalline Pr$_{2}$Ba$_{4}$Cu$_{7}$O$_{15-\delta }$, to examine some differences in the electronic state between  the reduced and as-sintered samples. We note that the $S$ behavior of both  as-sintered and reduced samples prepared by a sol-gel method\cite{TO12} is, in its magnitude and temperature dependence, very similar to the present data. 
Upon decreasing temperature, the value of $S$ of the superconducting sample reaches a broad maximum around $\sim  $230 K, then shows a rapid decrease below 200 K, and finally follows a substantial drop associated with the superconducting transition at low temperatures below $T_\mathrm{c}^\mathrm{on}=26.5 $K.  On the other hand,  $S$ of the non superconducting sample shows a stronger temperature dependence than that of the superconducting one at higher temperatures, and then approaches a very small value ($\sim 1\mu  $V/K) at the intermediate temperatures below 100 K.  These findings of the as-sintered Pr$_{2}$Ba$_{4}$Cu$_{7}$O$_{15-\delta }$ are in good agreement with the temperature variation of $S$ of PrBa$_{2}$Cu$_{4}$O$_{8 }$(Pr124) \cite{TE96} over the whole  range of temperature. In the quarter filling of the 1D CuO double chain  the static charge ordering  does not appear due to the interchain transfer and instead of it charge freezing is observed in NQR experiment of Pr124.\cite{FU03} We thus expect that the anomalous behavior of Seebeck coefficient for both the as-sintered Pr247  and  Pr124  is closely related to the presence of charge freezing. 
%The interchain transfer across CuO double chain suppresses insulating state associated with charge ordering.
The 1D upper and lower bands of CuO two single chains in each double chain are separated by charge gap at the absence of  the interchain transfer.  Because of weak interaction between the two single chains,   both band edges of the double chain reach Fermi energy and yield a small overlap in energy,  leading to the compensated half metallic state.  It seems that the value of $S$ oscillates along the horizontal axis upon lowering temperature  below 100 K, indicating the existence of two kind of carriers such as electrons and holes.   If in the half metal picture one lower band is almost filled, and another upper band is nearly empty, we then expect that holes  are almost  compensated with electrons. The present two band model also reproduces  the experimental result that Hall coefficient $R_\mathrm{H}$ of the as-sintered Pr247 below 100 K is, in its magnitude, much lower than that of the  electron doped  sample by reduction treatment.\cite{MA07}
  
%Furthermore, to further understand Seebeck coefficient, we show in Fig.\ref{S}(b) an energy dispersion relation and velocity derivative with respect to $k$, $E(k)$ and  $v'(k)$, for the $d$ band of the CuO double chain calculated by using the $d-p$ double chain model.\cite{SA05}
For conventional metal, Seebeck coefficient is expressed in such a way,

%\[S(T)=\frac{{(\pi k_{B})}^{2}T}{3e} \left\{ \frac{\partial \ln (N(E))}{\partial E}  \right\ } _{E=E_{F}},\]

\[S(T)= \frac{(\pi k_\mathrm{B})^{2}T}{3e} \left\{ \frac{d \ln ( N(E))}{d E}  \right\} _{E=E_\mathrm{F}},\]
\[= \frac{(\pi k_\mathrm{B})^{2}T}{3e} \left\{\frac{1}{N}  \frac{d N(E)}{d E}  \right\} _{E=E_\mathrm{F}},\]
%$S=(\pi k)^2T/3e\{\partial $ln$(N(E))/\partial E \}$=$(\pi k)^2T/3eN(E)
%\{\partial $$(N(E))/\partial E \}$
where $N(E)$ is the density of electronic state (DOS).\cite{ZI76} 
Here, we assume the energy dispersion $E(k)$ of the double chain model given in ref. \cite{HA11}.
The inter-chain hopping integrals are neglected since their contribution to the energy derivative of the DOS is  very small except for the band edges near $n\simeq 0$. 
By using the following formula and calculating the energy derivative of the DOS as a function of the electron density $n$,

%\[\frac{1}{N}\frac{d N}{d E}=-\frac{v'_{F_{1}}}{v^{3}_{F_{1}}}-\frac{v'_{F_{2}}}{v^{3}_{F_{2}}},\]

\[\frac{1}{N}\frac{d N}{d E}= \frac{1}{1/v_\mathrm{F_{1}}+1/v_\mathrm{F_{2}}}\left(-\frac{v'_\mathrm{F_{1}}}{v^{3}_\mathrm{F_{1}}}-\frac{v'_\mathrm{F_{2}}}{v^{3}_\mathrm{F_{2}}}\right) ,\]
the density functional band calculation then gives the calculated data as shown in Fig.\ref{S}(b).
Here, $v_\mathrm{F_{1}}$ and $v_\mathrm{F_{2}}$ represent two Fermi velocities in the double chain model, $v=dE/dk$, and $v'=dv/dk$.
From Fig.\ref{S}(b), the electron density up to $n=\sim 0.4$ gives rise to $S>0$ in the case of  $e<0$, which is consistent with the experimental result. Moreover, lower $S$ of the reduced sample  at high temperatures near 250 K  in comparison to that of the as sintered one supports a monotonous decrease of the absolute value of  $\frac{1}{N}\frac{dN}{dE}$ due to electron doping. 
The value $R_\mathrm{H}$ for the citrate sample measured at 20 K  is $-0.77\times 10^{-3}$ cm$^{3}$/C (not shown here), which is not so far from the data of Pr247 prepared by high pressure synthesis. 
The effect on magnetic field on Seebeck coefficient of the superconducting sample is qualitatively  similar to that on the resistivity as displayed in the inset of Fig.\ref{S}(a). The field dependence of  $S$ for the superconducting Pr247 up to 12 T  does not approach the behavior of the as-sintered non-superconducting $S$.This finding also supports a quite difference in the electronic state between the as-sintered and reduced samples in Pr247. 
%In our rough estimation, if we assume that the magnitude of S is proportinal to the energy derivative of  $N(E)$,  the density of state for 2D metal is independent of the energy, giving $S(T)\sim 0$.

\section{SUMMARY}
We have investigated the magnetotransport and $dc$ magnetic susceptibility  of the polycrystalline samples of Pr$_{2}$Ba$_{4}$Cu$_{7}$O$_{15-\delta }$, to examine the effect of magnetic field on the superconducting phase of the metallic CuO double chain.  
%A reduction treatment of the as-sintered sample in vacuum causes higher superconductivity achieving $T_{c,on}=\sim 30$ K for $\delta =0.94$.  
The resistive critical magnetic field is estimated to be about 21 T at low temperatures from the resistive transition data. On the other hand, the corresponding critical field determined from the magnetization measurements gives rise to a very low value of $\sim 0.3$T at 2 K. These discrepancies in the magnetic response between the resistivity and magnetization data are caused  by  disappearance  of the magnetically shielding effect even in relatively lower fields. In spite of the observation of the resistive drop associated with the superconducting transport currents, the suppression of the diamagnetic signal is probably related to the superconductivity of quasi one-dimensional CuO double-chain.  We report temperature dependence of Seebeck coefficient $S(T)$ of the polycrystalline Pr$_{2}$Ba$_{4}$Cu$_{7}$O$_{15-\delta}$ , to examine some differences in the electronic state between  the reduced and as-sintered samples. The behavior of  $S(T)$ in the superconducting Pr247 is discussed on the basis of the double chain model  from the density functional band calculation.

%X-ray powder diffraction pattern of nominal Pr$_{2}$Ba$_{4}$Cu$_{7}$O$_{15-\delta }$ reveals the coexistence of Pr123 and Pr124 phases.  
\begin{acknowledgments}
The authors are grateful for  M. Nakamura for his assistance in PPMS experiments at Center for Regional Collaboration in Research and Education, Iwate University. They thank Dr. A. Matsushita for his collaboration in Hall coefficient measurement. 
%This work was partially supported by a Grant-in-Aid for Scientific Research from Japan Society of the Promotion of Science. 
\end{acknowledgments}

\end{document}